\documentclass[aps,prl,reprint,superscriptaddress]{revtex4-2}
\usepackage{graphicx}
\usepackage{graphicx}
\usepackage{dcolumn}
\usepackage{bm}
\usepackage[dvipsnames]{xcolor}
\usepackage{amsmath}

\newcommand{\bfpsi}{\mbox{\boldmath$\psi$}}

\begin{document}

\title{Antiferromagnetic order in a layered magnetic topological insulator MnBi$_2$Se$_4$ probed by resonant soft x-ray scattering}

\author{Xiang Chen} 
\email{Current affiliation: Sun Yat-Sen University, Guangzhou, China; Email: chenx889@mail.sysu.edu.cn}
\affiliation{Materials Science Division, Lawrence Berkeley National Lab, Berkeley, California 94720, USA}
\affiliation{Physics Department, University of California, Berkeley, California 94720, USA}

\author{Alejandro Ruiz}
\affiliation{Department of Physics, University of California, San Diego, California 92093, USA}

\author{Alexander J. Bishop}
\affiliation{Department of Physics, The Ohio State University, Columbus, Ohio 43210, USA}

\author{Brandon Gunn}
\affiliation{Department of Physics, University of California, San Diego, California 92093, USA}

\author{Rourav Basak}
\affiliation{Department of Physics, University of California, San Diego, California 92093, USA}

\author{Tiancong Zhu} 
\affiliation{Physics Department, University of California, Berkeley, California 94720, USA}

\author{Yu He}
\affiliation{Department of Applied Physics, Yale University, New Haven, Connecticut, 06511, USA}
\affiliation{Physics Department, University of California, Berkeley, California 94720, USA}
\affiliation{Materials Science Division, Lawrence Berkeley National Lab, Berkeley, California 94720, USA}

\author{Mayia Vranas}
\affiliation{Department of Physics, University of California, San Diego, California 92093, USA}

\author{Eugen Weschke} 
\affiliation{Helmholtz-Zentrum Berlin für Materialen und Energie, BESSY II, D-12489 Berlin, Germany. }

\author{Roland K. Kawakami}
\affiliation{Department of Physics, The Ohio State University, Columbus, Ohio 43210, USA}

\author{Robert J. Birgeneau}
\affiliation{Physics Department, University of California, Berkeley, California 94720, USA}
\affiliation{Materials Science Division, Lawrence Berkeley National Lab, Berkeley, California 94720, USA}
\affiliation{Department of Materials Science and Engineering, University of California, Berkeley, California 94720, USA}

\author{Alex Frano}
\email{afrano@ucsd.edu }
\affiliation{Department of Physics, University of California, San Diego, California 92093, USA}

\date{\today}

\begin{abstract}

The quasi-two-dimensional magnetic topological insulator MnBi$_2$Se$_4$, stabilized via non-equilibrium molecular beam epitaxy, is investigated by resonant soft x-ray scattering. Kiessig fringes are observed, confirming a high sample quality and a thin film thickness of 10 septuple layers ($\sim$13 nm). An antiferromagnetic Bragg peak is observed at the structurally forbidden reflection, whose magnetic nature is validated by studying its temperature, energy, and polarization dependence. Through a detailed analysis, an A-type antiferromagetic order with in-plane moments is implied. This alternative spin structure in MnBi$_2$Se$_4$, in contrast to the Ising antiferromagnetic states in other magnetic topological insulators, might be relevant for hosting new topological states.

\end{abstract}

\maketitle


\section{Introduction}

The newly uncovered axion insulator states with quantized magnetoelectric effects, Chern insulator phases, and quantum anomalous Hall effect with dissipationless chiral edge states in magnetic topological insulators, open the door for potential applications in next-generation spintronics and quantum computing \cite{Haldane_1988_PRL_QHE, Essin_2009_PRL_Axion, Li_2010_NP_Axion, Mong_2010_PRB_ATI, Hasan_2010_RMP_TI, XL_Qi_2011_RMP_TI, Tokura_2019_NRP_MTI, Chang_2013_Science, Jiaheng_2019_SciAdv, Gong_2019_CPL_MTI, Otrokov_2019_Nature_MTI, Otrokov_2019_PRL, Zhang_2019_PRL_Topological_Axion, C_Liu_2020_NatMat_Axion, Deng_2020_Science, Bernevig_2022_Nature_MTI, Chang_2023_RMP}. The coupling of magnetism and topology is essential to support these exotic quantum topological states. The recently discovered quasi-two-dimensional (quasi-2D) van der Waals-bonded (vdW) compound MnBi$_2$Te$_4$ \cite{Gong_2019_CPL_MTI, Otrokov_2019_Nature_MTI, Otrokov_2019_PRL}, in which the intralayer manganese (Mn) atoms within a septuple layer (SL) (i.e., Te-Bi-Te-Mn-Te-Bi-Te) are magnetically ordered, provides important opportunities to explore these rich magnetic topological states. As a close analogue, the sister compound MnBi$_2$Se$_4$ is predicted to host either a nodal line with in-plane magnetic moments or a Weyl semimetal state with Ising spins  \cite{Chowdhury_2019_npjCM_MnBi2Se4, Zhang_2021_PRB_MnBi2Se4}, further highlighting the significance of magnetic interactions. Therefore, understanding the magnetic ground states in these materials is a prerequisite for further exploring new topological states.

The exploration of quantum topological states in the quasi-2D vdW magnet MnBi$_2$Se$_4$ is hindered by the extreme difficulty in single crystal synthesis. Efforts to synthesize bulk crystals of vdW MnBi$_2$Se$_4$ have only produced its monoclinic form \cite{Lee_1993_JAC, Nowka_2017_JCG_MnBi2Se4}, which is a thermodynamically more stable phase. By using non-equilibrium molecular beam epitaxy (MBE) to stabilize the vdW layers, multilayer vdW MnBi$_2$Se$_4$ crystal films were recently synthesized \cite{Zhu_2021_NL_MnBi2Se4, Walko_2022_PELSN_MnBi2Se4}. Although magnetization measurements of the vdW MnBi$_2$Se$_4$ thin films suggest the existence of ferromagnetism with an ordering temperature $\sim$10 K, as evidenced by the temperature-dependent magnetization and the out-of-plane hysteresis loops of the isothermal magnetization, an antiferromagnetic (AFM) order has also been implied \cite{Zhu_2021_NL_MnBi2Se4}. Due to the limited volume of thin films (up to 20 SLs), neutron scattering studies of the magnetic properties of MnBi$_2$Se$_4$ are difficult. Because of the aforementioned challenges, the magnetic ground state of layered MnBi$_2$Se$_4$ remains elusive.

Resonant (elastic) soft x-ray scattering (RSXS) \cite{Blume1985, Hannon1988PRL, Fink_2013_RPP, Comin_2016_ARCMP, Paolasini_2014_resonant} offers a unique, element-specific probe to study spatial modulations of the spin degrees of freedom in solids on a nanoscopic length scale. However, soft x-rays are subject to strong absorption by ambient atmosphere and the penetration depth into a solid, single-crystalline material is on the order of a few to hundreds of nanometers, which means that surface effects may become relevant. Therefore, RSXS setups are designed and maintained under high vacuum conditions, which pose challenges for investigating magnetic systems at very low temperatures (below 20 K). Here, we demonstrate the applicability of utilizing the RSXS technique to investigate the bulk magnetic properties of a 10-SL MnBi$_2$Se$_4$ thin film sample (thickness $\sim$13 nm). By tuning the x-ray energy $E=\hbar\omega$ to the Mn $L_3$ absorption edge, the photon-excited intermediate state is sensitive to the magnetic order via the dipole allowed Mn 2$p$ $\to$ 3$d$ ($L_{\text{3}}$) transition involved in the resonant elastic scattering process \cite{Blume1985, Hannon1988PRL}. Our study implies that the magnetic ground state of the quasi-2D vdW magnet MnBi$_2$Se$_4$ is an A-type AFM order below a N{\'e}el temperature $T_{\text{N}}$ $\approx$ 9 K.


\begin{figure}[t]
\centering
\includegraphics[width= 8.5 cm]{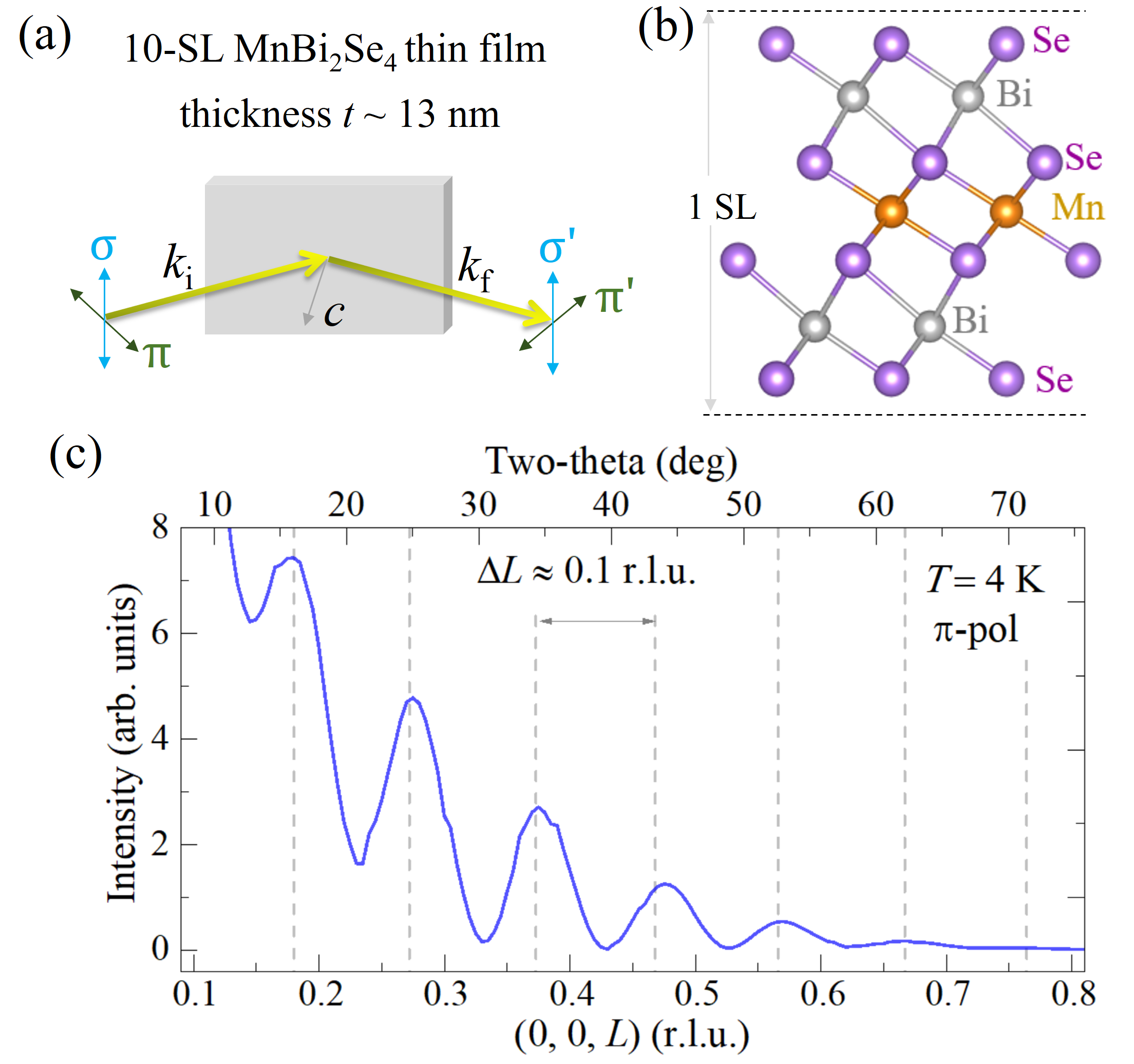}
\caption{(Color online) (a) Schematic of the resonant soft x-ray scattering (RSXS) experiment on a 10-septuple-layer (10-SL) MnBi$_2$Se$_4$ thin film with a thickness of $t$ $\approx$ 13 nm. The lattice $c$ direction lies within the horizontal scattering plane. The incoming x-ray can be either horizontally polarized ($H$-pol or $\pi$-pol) or vertically polarized ($V$-pol or $\sigma$-pol). (b) The atomic structure of one SL MnBi$_2$Se$_4$. Colored atoms: Se (purple), Bi (gray), Mn (orange). (c) Longitudinal $L$ scan of the 10-SL MnBi$_2$Se$_4$ thin film at $T$ = 4 K. Kiessig fringes \cite{Decher_1997_Science_Kiessig, Katmis_2016_Nature_Kiessig} are evident with an average periodicity of $\Delta L$ = 0.1 r.l.u., which is consistent with the sample thickness $t$ $\approx$ 13 nm. The corresponding two-theta angles are also labelled.}
\label{fig:Fig1_alpha}
\end{figure}


\section{Experimental results}

Single-crystalline vdW MnBi$_2$Se$_4$ thin films were synthesized layer-by-layer using the MBE technique under non-equilibrium condition \cite{Zhu_2021_NL_MnBi2Se4}. A 10-SL thick MnBi$_2$Se$_4$ film on Al$_2$O$_3$ (0001) with a $\sim$5 nm selenium (Se) cap was prepared. The RSXS experiments were performed at the UE46$_{-}$PGM-1 beamline at Helmholtz-Zentrum Berlin. A horizontal scattering geometry was utilized with the sample lattice $c$ direction lying within the scattering plane [Fig. 1(a)] \cite{Fink_2013_RPP, Chen_2022_FeCo_PRM}. The Bragg peaks $\textbf{Q}$ = $(H, K, L) \cdot (\frac{4\pi}{\sqrt{3}a}, \frac{4\pi}{\sqrt{3}b}, \frac{2\pi}{c})$ are defined in reciprocal lattice units (r.l.u., the latter bracket) with lattice parameters $a$ = $b$ $\approx$ 4.0 \AA \, and $c$ $\approx$ 12.8 \AA \, \cite{Zhu_2021_NL_MnBi2Se4}. Here for simplicity, the lattice constant $c$ is defined as one SL thickness $c_{\text{SL}}$ [Fig. 1(b)], instead of 3$c_{\text{SL}}$ as in MnBi$_2$Te$_4$. The x-ray scattering data are collected near the Mn $L_3$ edge to enhance the magnetic scattering signal \cite{Wilkins_2003_PRL_LaSrMnO, Windsor_2014_PRL_LuMnO3_RSXS, Padmanabhan_2022_AM_MnBi2Se4_RSXS}. The incoming x-rays can be either horizontally polarized ($H$-pol or $\pi$-pol) or vertically polarized ($V$-pol or $\sigma$-pol), but the polarization of the scattered x-rays is not analyzed. Therefore, both outgoing $\sigma'$-pol and $\pi'$-pol channels contribute to the scattered intensity. 

In general for a resonant x-ray scattering experiment, the scattering amplitude consists of both charge and magnetic scattering, and can be written as \cite{Blume1985, Hannon1988PRL, Paolasini_2014_resonant}:

\begin{equation}
f = f_0 (\textbf{e}_{\nu}^{\ast} \cdot \textbf{e}_{\mu}) - \text{i} f_1 ( \textbf{e}_{\nu}^{\ast} \times \textbf{e}_{\mu}) \cdot \textbf{m} + f_2 (\textbf{e}_{\nu}^{\ast} \cdot \textbf{m}) (\textbf{e}_{\mu} \cdot \textbf{m})
\label{eq:eqn1}
\end{equation}

in which, $f_0$, $f_1$ and $f_2$ are the monopole, magnetic dipole ($E_1$) and quadrupole ($E_2$) part of the energy dependent resonance amplitude, respectively. $\textbf{e}_{\mu}$ ($\mu = \sigma$ or $\pi$) and $\textbf{e}_{\nu}$ ($\nu = \sigma'$ or $\pi'$) are unit vectors along the polarization direction of the electric field component of the incident and out-going x-ray beams, respectively. $\textbf{m}$ is the unit vector along the spin direction. At the 3$d$ transition metal $L$ edges, the $E_1$ resonant scattering is often greatly enhanced. Meanwhile, by tuning the x-ray polarization and energy, as well as the scattering geometry, the charge scattering part can sometimes be much reduced, such that considering only the magnetic dipole term in equation (1) is sufficient for describing the resonant x-ray scattering intensity \cite{ Paolasini_2014_resonant, Biffin_2014_PRB_Li2IrO3, Boseggia_2013_JPCM, Chen_2022_FeCo_PRM}. Thus, because of the large resonant magnetic scattering amplitude at the Mn $L$ edges and the inversion symmetry at the Mn sites, the resonant magnetic scattering intensity at a structurally forbidden Bragg peak position can be approximated by including the linear resonant scattering amplitude in magnetic moment (without considering the charge scattering amplitude) \cite{Fink_2013_RPP, Comin_2016_ARCMP, Paolasini_2014_resonant, Biffin_2014_PRB_Li2IrO3, Boseggia_2013_JPCM}:

\begin{equation}
I_{\mu \nu} \propto | \sum_{j} e^{ i \textbf{Q} \cdot \textbf{r}_j} ( \textbf{e}_{\mu} \times \textbf{e}_{\nu}^{\ast}) \cdot \textbf{m}_j F(\text{E})|^2,
\label{eq:eqn2}
\end{equation}

to first order in the magnetic moment $\textbf{m}_j$ of the ion located at site $\textbf{r}_j$ within the unit cell.  $F(\text{E})$ is the non-local, photon energy dependent scattering tensor. Since the polarization of the scattered light is not analyzed, the measured intensities are $I_{\text{H}}$ = $I_{\pi}$ $\equiv$ $I_{\pi \sigma'}$ + $I_{\pi \pi'}$ and $I_{\text{V}}$ = $I_{\sigma}$ $\equiv$ $I_{\sigma \sigma'}$ + $I_{\sigma \pi'}$. Also, since $\textbf{e}_{\sigma}$ $||$ $\textbf{e}_{\sigma'}$, this obviously leads to $I_{\sigma \sigma'}$ = 0, and leaves $I_{\sigma}$ = $I_{\sigma \pi'}$.

\begin{figure}[t]
\centering
\includegraphics[width = 6.5 cm]{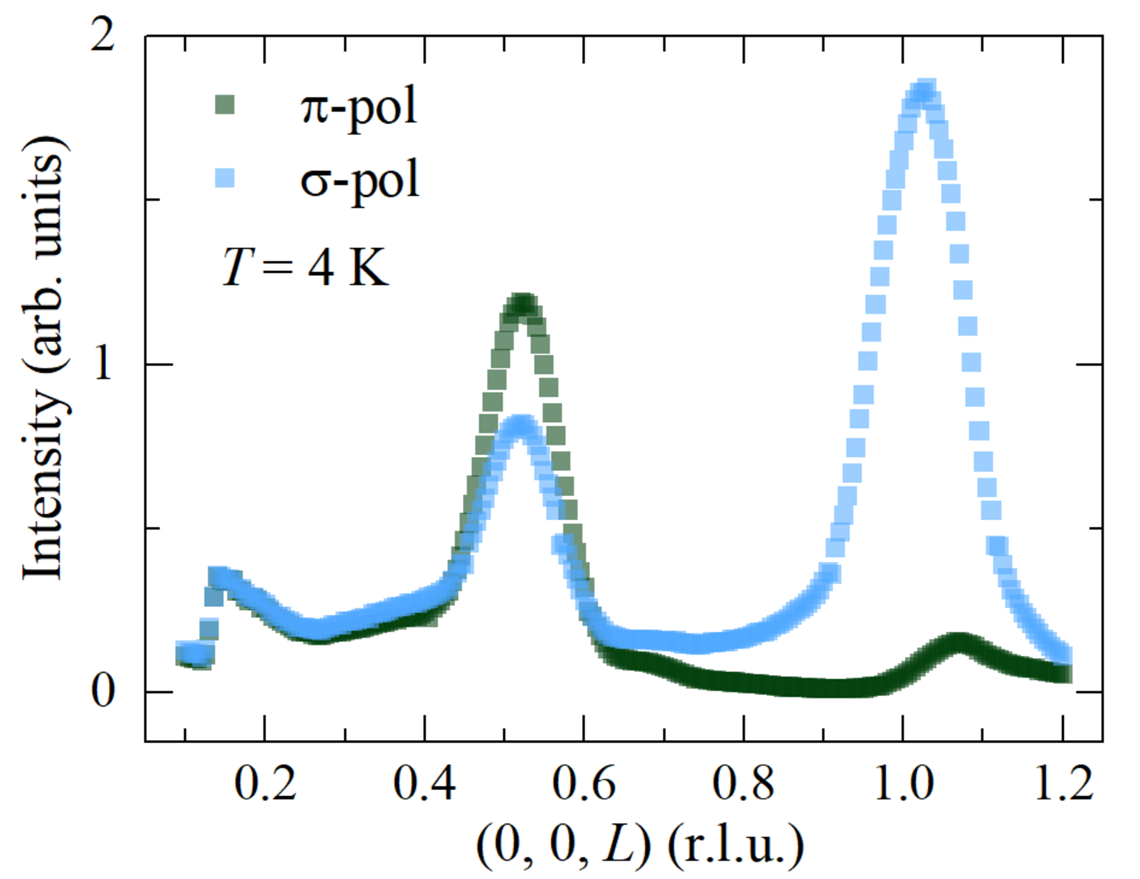}
\caption{(Color online) Comparison of the longitudinal $L$ scan at the Mn $L_{\text{3}}$ edge (energy $E$ = 640.2 eV) at $T$ = 4 K with two different polarizations of incoming x-rays: $\pi$-pol (green) and $\sigma$-pol (blue). With $\pi$-pol incoming photons, the structural Bragg peak (0, 0, 1) is greatly suppressed; while the intensity of the magnetic Bragg peak (0, 0, 0.5) is only moderately enhanced from $\sigma$-pol to $\pi$-pol. The intensity jump near $L$ = 0.15 r.l.u. is an artifact.}
\label{fig:Fig2_alpha}
\end{figure}

The quasi-2D vdW magnet MnBi$_2$Se$_4$ is isostructural to MnBi$_2$Te$_4$ (space group $R\bar{3}m$), with covalently bonded SLs (i.e., Se-Bi-Se-Mn-Se-Bi-Se) as the base unit [Fig. 1(b)]. The atomic-scale structures of MnBi$_2$Se$_4$ were verified via scanning transmission electron microscopy, scanning tunneling microscopy, and atomic force microscopy, confirming the layered vdW structure, as oppose to the monoclinic form produced by single-crystal synthesis \cite{Zhu_2021_NL_MnBi2Se4, Walko_2022_PELSN_MnBi2Se4}. The soft x-ray scattering experiments were performed on a 10-SL MnBi$_2$Se$_4$ thin film sample. From the longitudinal $L$ scan of the x-ray diffraction measurement [Fig. 1(c)], well defined Kiessig fringes \cite{Decher_1997_Science_Kiessig, Katmis_2016_Nature_Kiessig} are evident with an average periodicity of $\Delta L$ = 0.1 r.l.u., indicating a coherent interference of the x-ray beams reflected on the top cap layer of amorphous Se and the interface of MnBi$_2$Se$_4$ and Al$_2$O$_3$ (0001). The large oscillating amplitude of the Kiessig fringes at a low two-theta angle, as compared to the peak height of the charge/magnetic peak in Figs 2-3, implies a small surface roughness of the films. Moreover, the thickness of the MnBi$_2$Se$_4$ film can be readily estimated as $t$ = $\frac{2\pi}{\Delta L}$ = 10 $c_{\text{SL}}$ $\approx$ 13 nm, which matches in excellent agreement with the targeted MBE film growth thickness of 10 SLs.


\begin{figure}[t]
\centering
\includegraphics[width= 7.0 cm]{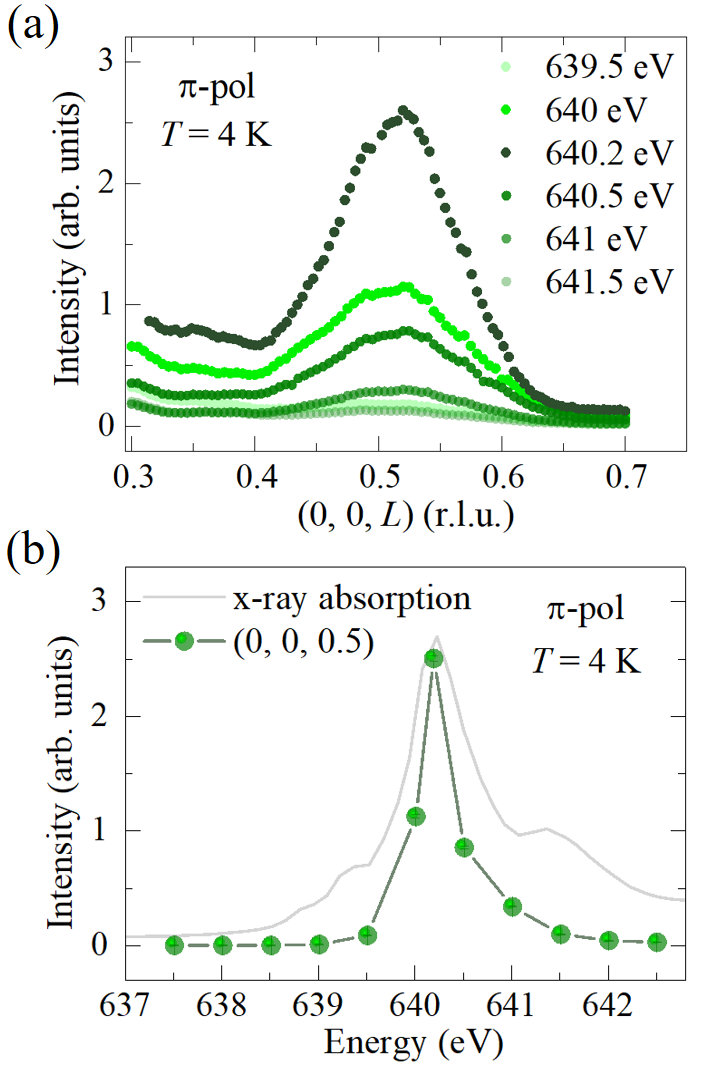}
\caption{(Color online) Energy dependence of the (0, 0, 0.5) peak at $T$ = 4 K with $\pi$-pol incoming x-rays: (a) $L$ scans of the (0, 0, 0.5) peak at select energies near the Mn $L_{\text{3}}$ edge. (b) Energy dependence of the integrated area of the $L$ scans of the (0, 0, 0.5) peak (green circles). For comparison, the x-ray absorption data are also plotted (solid gray line).}
\label{fig:Fig3_alpha}
\end{figure}

\begin{figure}[t]
\centering
\includegraphics[width= 8.5 cm]{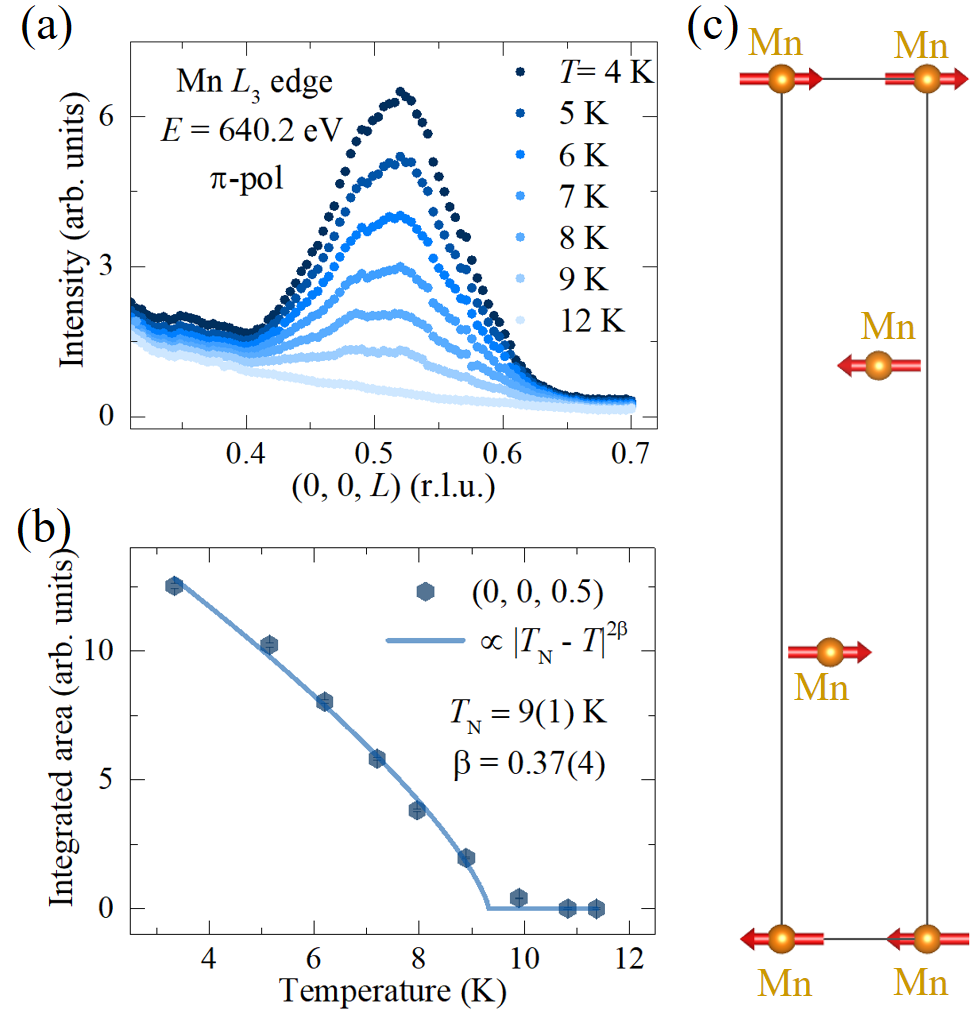}
\caption{(Color online) Temperature dependence of the (0, 0, 0.5) peak on resonance (at $E$ = 640.2 eV), collected with $\pi$-pol incoming x-rays: (a) $L$ scans of the (0, 0, 0.5) peak at select temperatures. The peak intensity is completely suppressed at $T$ = 12 K. (b) Temperature dependence of the integrated area of the (0, 0, 0.5) peak. The solid blue line is a power law fit to the data with a form: $I \propto (T_{\text{N}} - T)^{2\beta}$, which yields $T_{\text{N}}$ = 9(1) K and $\beta$ = 0.37(4). (c) A-type antiferromagnetic order in MnBi$_2$Se$_4$ with in-plane Mn moments. For simplicity, only Mn atoms within the unit cell of 3 SLs are shown.}
\label{fig:Fig4_alpha}
\end{figure}


By tuning the x-ray energy to the Mn $L_3$ absorption edge ($E$ $\approx$ 640 eV) and slightly tilting the rocking angle off by 0.25$^{\circ}$ (Figs. S1-S2, \cite{suppfile}), which suppress the Kiessig fringes without affecting the magnetic signals, a Bragg peak reveals at the structurally forbidden position $\textbf{Q}_0$ = (0, 0, 0.5) from the longitudinal $L$ scan below the magnetic onset temperature (at $T$ = 4 K), in addition to the structural Bragg peak at $\textbf{Q}_1$ = (0, 0, 1) (Fig. 2). The contrasting nature of the Bragg peaks $\textbf{Q}_0$ and $\textbf{Q}_1$ is apparent; opposing polarization dependent behavior is clearly observed when the polarization of the incoming x-rays is switched from $\sigma$-pol to $\pi$-pol (Fig. 2). The structural Bragg peak intensity at $\textbf{Q}_1$ is greatly reduced with $\pi$-pol incoming x-rays; while the magnetic Bragg peak $\textbf{Q}_0$ is only moderately influenced by varying the polarization of the incoming x-rays (Fig. 2). As discussed later, the stronger signal at $\textbf{Q}_0$ for $\pi$-pol is consistent with Mn moments oriented in-plane as opposed to out-of-plane, as observed in magnetization measurements \cite{Zhu_2021_NL_MnBi2Se4}.

To understand fully the nature of the $\textbf{Q}_0$ peak, we present a further study of its dependence on x-ray energy and temperature. Fig. 3 shows the photon energy dependence of the $\textbf{Q}_0$ reflection, collected at 4 K with $\pi$-pol incident x-rays. The $\textbf{Q}_0$ peak intensity shows a dramatic dependence on the photon energy. A single and sharp energy-dependent peak profile with an estimated peak width of $\sim$0.6 eV is evident near the Mn $L_3$ edge, in which the maximum intensity of the $\textbf{Q}_0$ reflection is recorded at $E$ = 640.2 eV [green circles in Fig. 3(b)]. As a comparison, the x-ray absorption spectra at the Mn $L_3$-edge is also plotted [gray line in Fig. 3(b)]. This sharp resonance in photon energy is a strong signature of the magnetic nature of the $\textbf{Q}_0$ reflection \cite{Fink_2013_RPP, Comin_2016_ARCMP} because the resonant scattering proceeds in two stages, occurs in conjunction with an absorption edge, and the scattered intensity depends on both the incoming and outgoing photon polarizations.

The temperature dependence of the $\textbf{Q}_0$ peak at the resonance energy ($E$ = 640.2 eV) is also examined, as shown in Fig. 4. With increasing temperature, the $\textbf{Q}_0$ peak intensity on resonance becomes vanishingly small above the magnetic transition temperature $T_{\text{N}}$ $\approx$ 10 K [Figs. 4(a)-(b)]. We argue that, from the polarization-, temperature-, and energy-dependent studies, the $\textbf{Q}_0$ = (0, 0, 0.5) Bragg peak in Figs. 2-4 is dominantly magnetic in nature; while the charge scattering contribution is greatly suppressed by tuning the polarization and the energy of the incident x-rays, as well as tilting the rocking angles (Figs. S1-S2, \cite{suppfile}). This magnetic (0, 0, 0.5) Bragg peak implies the AFM nature of the magnetic ground state in vdW MnBi$_2$Se$_4$.

To gain more insight about the nature of the magnetic transition in MnBi$_2$Se$_4$, a power law fitting of the order parameter squared \cite{Guida_1998_JPAMG,Pelissetto2002,Birgeneau1977_PRB,Chen_2021_PRB_Ba4310}---the integrated peak intensity after subtracting the high-temperature background at $T$ = 12 K---is employed at the magnetic reflection $\textbf{Q}_0$ = (0, 0, 0.5) under the form:

\begin{equation}
I \propto | T_{\text{N}} - T |^{2\beta} , 
\label{eq:eqn4}
\end{equation}

The power law fit of the $\textbf{Q}_0$ = (0, 0, 0.5) magnetic peak yields $T_{\text{N}}$ = 9(1) K and $\beta$ = 0.37(4) [Fig. 4(b)]. The fitted value $T_{\text{N}}$ = 9(1) K, reflecting a bulk magnetic transition, is consistent with the temperature estimated from magnetization data. It is worth mentioning that the $\beta$ = 0.37(4) value is consistent, within the range of errors, with that of MnBi$_2$Te$_4$, which is $\beta$ = 0.32(1) or 0.35(2), determined from neutron diffraction experiments \cite{Ding_2020_PRB_MnBi2Te4, JQ_Yan_2019_PRM_MnBi2Te4}.

The ordered spin moment direction of the Mn ions in vdW MnBi$_2$Se$_4$ below $T_{\text{N}}$ can be inferred from representational analysis \cite{Wills2000}, the photon polarization dependence, as well as the calculation of the magnetic structure factor from the RSXS data. The magnetic representation of the crystallographic site of Mn in vdW MnBi$_2$Se$_4$ can be decomposed in terms of the irreducible representations (IRs) \cite{Wills2000}, with a propagation vector (0, 0, 1.5) (Table I) \cite{supp}: $\Gamma_{\text{Mag}}= 1\Gamma_{3}^{1}+1\Gamma_{5}^{2}$, where $\Gamma_{3}$ is a one-dimensional IR with magnetic moments pointing parallel to the $c$ axis (magnetic space group $R_{I}\bar{3}c$) and $\Gamma_{5}$ is a two-dimensional IR with basis vectors lying in the $ab$-plane (magnetic space group $C_{c}2/c$ or $C_{c}2/m$). The solution from $\Gamma_{3}$ can be excluded from the following reasoning.

Experimentally, the magnetic peak intensity at the $\textbf{Q}_0$ peak position with $\pi$-pol incoming x-rays is larger than that of $\sigma$-pol x-rays (Fig. 2). This means $I_{\pi}$ $>$ $I_{\sigma}$, which can be more specifically written as:

\begin{equation}
 \begin{split}
   |  \sum_{j} e^{ i \textbf{Q}_0 \cdot \textbf{r}_j}  \textbf{k}_i \cdot \textbf{m}_j  |^2 + |  \sum_{j} e^{ i \textbf{Q}_0 \cdot \textbf{r}_j}  \textbf{e}_{\sigma} \cdot \textbf{m}_j  |^2  \\
   >   |  \sum_{j} e^{ i \textbf{Q}_0 \cdot \textbf{r}_j}  \textbf{k}_f \cdot \textbf{m}_j  |^2 .
 \end{split}
\label{eq:eqn2}
\end{equation}

where $\textbf{k}_i$ and $\textbf{k}_f$ are the unit vectors along the incoming and outgoing photon wave-vector directions, respectively. If the magnetic moments are pointing along the lattice $c$ direction, then for any given $\textbf{m}_j$, $\textbf{e}_{\sigma} \cdot \textbf{m}_j$ = 0 and $\textbf{k}_i \cdot \textbf{m}_j $=$\textbf{k}_f \cdot \textbf{m}_j $, because $\textbf{m}_j$ $||$ $c$ is perpendicular to the $\textbf{e}_{\sigma}$ direction and bisects the $-\textbf{k}_i$ and $\textbf{k}_f$ vectors. This directly leads to $I_{\pi}$ $=$ $I_{\sigma}$. Therefore, the experimentally established $I_{\pi}$ $>$ $I_{\sigma}$ does not support an Ising moment direction in vdW MnBi$_2$Se$_4$. A similar analysis can be applied to the in-plane spin configuration, which results in $I_{\pi \sigma'}$ =  $I_{\sigma \pi'}$ = $I_{\sigma}$. In this case, $I_{\pi \pi'}$ is typically nonzero because $\textbf{e}_{\sigma} \cdot \textbf{m}_j$ $\ne$ 0 (unless in the unique configuration $\textbf{e}_{\sigma}$ $||$ $\textbf{m}_j$) and the magnetic domains equivalently rotated by $120^\circ$ coexist. Therefore, $I_{\pi}$ $=$ $I_{\pi \sigma'}$ + $I_{\pi \pi'}$ $>$ $I_{\sigma}$ is naturally satisfied with the in-plane moment scenario. 

Alternatively, by assuming an out-of-plane spin moment direction, the magnetic structure factor at any Bragg peak reflection along the $L$ direction will be zero, which is exactly the case in the vdW material MnBi$_2$Te$_4$ \cite{Ding_2020_PRB_MnBi2Te4, JQ_Yan_2019_PRM_MnBi2Te4}. From the above arguments, including the analysis of the magnetic symmetry and the RSXS data, an A-type AFM order with an easy-plane anisotropy in vdW MnBi$_2$Se$_4$ is established, as depicted in Fig. 4(c). This is consistent with the magnetization measurements.

\begin{table}[t]
\begin{tabular}{ccc|cccccc}
  IR  &  BV  &  Atom & \multicolumn{6}{c}{BV components}\\
      &      &             &$m_{\|a}$ & $m_{\|b}$ & $m_{\|c}$ &$im_{\|a}$ & $im_{\|b}$ & $im_{\|c}$ \\
\hline
$\Gamma_{3}$ & $\bfpsi_{1}$ &      1 &      0 &      0 &     1 &      0 &      0 &      0  \\
$\Gamma_{5}$ & $\bfpsi_{2}$ &      1 &      0 &      1 &     0 &      0 &      0 &      0  \\
             & $\bfpsi_{3}$ &      1 &      2 &      1 &     0 &      0 &      0 &      0  \\
\end{tabular}
\caption{Magnetic symmetry analysis of the vdW MnBi$_2$Se$_4$. Basis vectors (BVs) for the space group R$\bar{3}$m with the propagation vector (0, 0, 1.5). The decomposition of the magnetic representation for the Mn site is 
$\Gamma_{\text{Mag}}= 1\Gamma_{3}^{1}+1\Gamma_{5}^{2}$. }
\label{basis_vector_table_1}
\end{table}


\section{Discussion}

Our work implies the AFM coupled nature in a 13 nm thick vdW MnBi$_2$Se$_4$ thin film below $T_{\text{N}}$ = 9(1) K, with intralayer ferromagnetic couplings and interlayer antiferromagnetic couplings. This is in good agreement with density functional theory (DFT) calculations for layered MnBi$_2$Se$_4$ \cite{Chowdhury_2019_npjCM_MnBi2Se4, Zhang_2021_PRB_MnBi2Se4}. The reported hysteresis behavior from the previous magnetization measurement is likely resulting from uncompensated layers \cite{Zhu_2021_NL_MnBi2Se4}, which display ferromagnetism due to the strong intralayer ferromagnetic exchange couplings. In addition, the scattering study of the MnBi$_2$Se$_4$ thin film disproves claims that the observed magnetism is simply due to some Mn atoms randomly doping the Bi$_2$Se$_3$ thin films, which give rise to weak ferromagnetism \cite{Zhang_2012_PRB_Mn_doped_Bi2Se3, Bardeleben_2013_PRB_Mn_doped_Bi2Se3}.


It should be noted that the in-plane anisotropy established in layered MnBi$_2$Se$_4$ by RSXS is not fully consistent with the DFT calculation, which predicts an Ising moment direction \cite{Chowdhury_2019_npjCM_MnBi2Se4, Zhang_2021_PRB_MnBi2Se4}. This discrepancy can be resolved by considering both the ligand $p$ orbital spin-orbit coupling effect and the magnetic dipole-dipole interactions \cite{Lado_2017_2DM, Kim_2019_PRL,  Xue_2019_PRB_CrCl3_PMA, Zhu_2021_NL_MnBi2Se4}. The Mn$^{2+}$ spin state in MnBi$_2$Te$_4$ is confirmed from the neutron diffraction measurements, which indicate an ordered magnetic moment 4.9(1) $\mu_{\text{B}}$/Mn and a $S$ = $\frac{5}{2}$ magnetic state with nearly quenched orbital angular momentum ($L \approx 0$) \cite{Ding_2021_JPAP}. However, magnetic exchange anisotropy can be induced by the ligand $p$ spin-orbit coupling through the super-exchange mechanism \cite{Lado_2017_2DM, Kim_2019_PRL}. Because of the increased atomic number from Se to Te, a larger ligand spin-orbit coupling is expected, resulting from the further extension of the Te $p$ orbitals. This explains the in-plane metallic behavior and a larger $T_{\text{N}}$ $\approx$ 24 K in MnBi$_2$Te$_4$ \cite{Otrokov_2019_Nature_MTI}, while more insulating behavior \cite{Zhu_2021_NL_MnBi2Se4} and a smaller $T_{\text{N}}$ $\approx$ 9 K are observed in MnBi$_2$Se$_4$ thin film samples. Therefore, a weakened perpendicular magnetization is expected in MnBi$_2$Se$_4$, compared to in MnBi$_2$Te$_4$. On the other hand, the magnetic shape anisotropy resulting from the magnetic dipole-dipole interactions usually supports an in-plane moment configuration. Consequently, because of the smaller magnetic exchange anisotropy resulting from the weaker super-exchange couplings and the magnetic shape anisotropy energy scale in layered MnBi$_2$Se$_4$, the spins are lying within the $ab$-plane, instead of the predicted Ising moment picture. 

It is worth mentioning that our heuristic model for the spin direction requires a more quantitative theoretical calculation to underpin the solution between $C_{c}2/c$ and $C_{c}2/m$. Nevertheless, the in-plane spin moment direction in vdW MnBi$_2$Se$_4$, in contrast to the Ising moments in other magnetic topological insulators such as MnBi$_2$Te$_4$, MnSb$_2$Te$_4$ and Mn$_2$Bi$_2$Te$_5$ \cite{Jiaheng_2019_SciAdv, Gong_2019_CPL_MTI, Otrokov_2019_Nature_MTI, Otrokov_2019_PRL, Chen_2019_NC_Sb_MnBi2Te4, Murakami_2019_PRB_MnSb2Te4, Ding_2021_JPAP}, provides an alternative route towards novel topological states (including Weyl line nodes) related to symmetry breaking by magnetic order \cite{Chowdhury_2019_npjCM_MnBi2Se4, Bernevig_2022_Nature_MTI, Li_2023_NL_MnBi2Se4}.

\section{Conclusion}

In summary, we successfully investigate the intrinsic magnetism in the quasi-2D vdW magnetic topological insulator MnBi$_2$Se$_4$ stabilized via the MBE technique under non-equilibrium condition. Through a resonant soft x-ray scattering study, an A-type antiferromagnetic order is demonstrated in a 10-SL MnBi$_2$Se$_4$ thin film, with a N{\'e}el temperature $T_{\text{N}}$ = 9(1) K. The thickness of the sample ($\sim$13 nm) is nicely determined by the Kiessig fringes. The magnetic order in layered MnBi$_2$Se$_4$ is confirmed through the temperature, energy, and polarization dependence of the (0, 0, 0.5) Bragg peak. In addition, the in-plane magnetic anisotropy is readily confirmed by the magnetic symmetry analysis and the polarization dependent study of the magnetic Bragg peak. Compared to other magnetic topological insulators with Ising moments, the alternative spin structure in vdW MnBi$_2$Se$_4$ might be useful for realizing new topological states.

\bigbreak

\section{Acknowledgments}

Work at the University of California, Berkeley and the Lawrence Berkeley National Laboratory was funded by the U.S. Department of Energy, Office of Science, Office of Basic Energy Sciences, Materials Sciences and Engineering Division under Contract No. DE-AC02-05-CH11231 within the Quantum Materials Program (KC2202). Work at UC San Diego was supported by the National Science Foundation under Grant No. DMR-2145080. Work at Ohio State was funded by AFOSR MURI 2D MAGIC Grant No. FA9550-19-1-0390 and
U.S. Department of Energy, Office of Science, Basic Energy Sciences Grant No. DE-SC0016379. We thank HZB for the allocation of synchrotron radiation beamtime. A.F. was supported by the Research Corporation for Science Advancement via the Cottrell Scholar Award (27551) and the CIFAR Azrieli Global Scholars program.

\bibliography{MnBiSe_main}

\end{document}